\documentclass[12pt]{article}
\usepackage{graphicx}
\usepackage{caption}
\usepackage{subcaption}
\usepackage{latexsym,graphicx,multirow}
\usepackage{float}
\usepackage{amssymb}
\usepackage{amsmath}
\usepackage{amscd}
\usepackage{amsthm}
\usepackage[left=2cm,top=2.5cm,right=2.5cm,bottom=1.5cm]{geometry}
\usepackage{hyperref}
\usepackage{epstopdf}
\usepackage{cite}
\usepackage{caption}
\usepackage{subcaption}
\usepackage{color}
\begin{document}
\begin{center}
	\large{\bf{Cosmological model in the framework of $f(R,\mathcal{L}_{m})$ gravity with quadratic equation of state parameter. }} \\
	\vspace{10mm}
	\normalsize{Vinod Kumar Bhardwaj$^{1,a}$, Saibal Ray$^{2,b}$ }\\
	\vspace{5mm}
	\normalsize{$^{1}$ Department of Mathematics, GLA University, Mathura-281 406, Uttar Pradesh, India}\\
	\vspace{2mm}
	\normalsize{$^{2}$ Centre for Cosmology, Astrophysics and Space Science (CCASS), GLA University, Mathura 281406, Uttar Pradesh, India}\\
	\vspace{2mm}
	%\normalsize{$^{3}$Centre of Cosmology Astrophysics and Space Science (CCASS), GLA University, Mathura-281 406, Uttar Pradesh, India}\\
	%\vspace{2mm}	
	$^a$E-mail:dr.vinodbhardwaj@gmail.com\\
	\vspace{2mm}
	$^b$E-mail:saibal.ray@gla.ac.in\\
	%\vspace{2mm}
	\vspace{10mm}	
	%\date{}
	%\maketitle
\end{center}	

\begin{abstract}
In this study, we have explored a transitioning cosmological model of universe's expansion in $f(R,\mathcal{L}_{m})$ gravity. The quadratic type of equation of state parameter in the form $\omega=-1 + \alpha (1 + z) + \beta (1 + z)^2$, where $\alpha$ and $\beta$ are constants, is considered to determine the explicit solution of field equations and derive Hubble parameter in term of redshift $z$. The model parameters are estimated taking observational datasets of BAO, Pantheon, and CC using MCMC analysis. Some dynamical properties like EOS parameter, energy density, pressure, and deceleration parameter are described. The cosmographic parameter like statefinders ($r,s$), jerk parameter ($j$) etc are also thoroughly explained. The energy conditions  are also examined to validate the viability of the proposed model. We observe a transition redshift at $z_{t} = 0.942^{+0.112}_{-0.164}$ with the present value of deceleration parameter $q_0 = -0.4815^{+0.0362}_{-0.0096}$. 
\end{abstract}
\smallskip 
{\bf Keywords} : $f(R,\mathcal{L}_{m})$ gravity, flat FLRW space-time, EOS parameter, Observational constraints, Energy conditions. 
%\maketitle
\section{Introduction} \label{sec1}

In the early 20th century, Einstein formulated the general relativity (GR) theory to illustrate the relationship between gravity, space, and time. He showcased a direct connection between matter, radiation, and the geometry of space-time. Linking energy and momentum, fundamental characteristics of matter and radiation with the curvature of space-time, Einstein developed the field equation expressed as $R_{ij}-\frac{1}{2}R g_{ij} =- 8\pi G T_{ij}$ \cite{ref1}. The cosmological models that describe the current state of the expanding universe are originated on the theory of general relativity. The current accelerated phase of the expanding universe is further confirmed by various experimental research \cite{ref2,ref3,ref4,ref5}. Dark energy (DE) is anticipated to be the key factor in the universe's expansion, whereas dark matter is believed to be the essential component for the formation of large-scale structures (LSS) \cite{ref6,ref7,ref8}. Due to its repulsive nature, the cosmological constant $\Lambda$ has been assumed to be an appropriate substitute for DE \cite{ref9}. Several cosmological models have been developed to describe current cosmic expansion, taking $\Lambda$ as the dark energy candidate \cite{ref9,ref10,ref11,ref12}. Although the standard $\Lambda$CDM model shows nice agreement with recent cosmological observations, it still faces the coincidence issue and fine-tuning problem on the theoretical ground \cite{ref13}. Apart from these issues, the $\Lambda$CDM model also suffers from $H_0$ tension. In order to resolve these issues and to clarify the origin and characteristics of dark energy, numerous dynamical dark energy models such as tachyons \cite{ref6}, phantom \cite{ref7,Phantom2024}, k-essence \cite{ref8,K-essence2023a,K-essence2023b,K-essence2023c,K-essence2024}, Chaplygin gas \cite{ref9,ref10,Chaplygin2010,Chaplygin2023,Chaplygin2024}, chameleon \cite{ref11}, quintessence \cite{ref12,ref13}, modified gravity frameworks \cite{ref14,ref15}, models incorporating extra dimensions \cite{ref16,ref17}, and scalar field theories  \cite{ref18,ref19,ref20}, have been suggested in the literature for explaining the present scenario of the universe's expansion.

By modifying Einstein's general relativity equations of motion, various alternate theories have been introduced to resolve the cosmological constant issues and explain the late-time cosmic accelerated expansion. The $f(R)$ is the most basic model that explored Einstein's (GR) theory by incorporating the arbitrary function $f(R)$ of the Ricci scalar $R$ within the gravitational action \cite{ref21,ref22}. The $ f(R) $ theory describes more advanced scenarios in comparison with GR and offers extensive measures \cite{ref23,ref24,ref25,ref26}. Starobinsky \cite{ref27} proposed a singularity-free isotropic cosmological model and Sotiriou \cite{ref28} gives scalar-tensor theory in the framework of $f(R)$ gravity. Nojiri and Odintsov \cite{ref29} reconstruct the modified theories to get the cosmological behavior of the universe and analyze Big Rip along with the future singularities. By modifying Einstein-Hilbert action's geometrical part, Harko et al. \cite{ref30} 2011 introduced another version of the modified theory termed $f(R, T)$ theory. Here, the matter Lagrangian is assumed as the function of the Ricci scalar and the trace of energy-momentum tensor ($T$).  One crucial idea of the $f(R, T)$ theory is that it leads to the possibility of particle generation by taking the quantum effect into account. Such scenarios are critical for astrophysical research because they suggest a connection between quantum and $f(R,T)$ theories. A few interesting works in different frame-works under modified $f(T)$ and $f(R,T)$ theory can be seen in refs. \cite{R5,R1,ref31,ref32,ref33,Deb2019a,Deb2019b,Maurya2021,ref34}.

Bertolami et al. \cite{ref35} suggest a modified version of the $f(R)$ gravity theory which involves directly connecting the matter Lagrangian density $L_{m}$ with a general function $f(R)$. Due to this connection between matter and geometry, an additional force perpendicular to the four velocity vectors arises when massive particles move off their geodesic paths. This model was expanded to include arbitrary connections in both substance and structure \cite{ref36}. Extensive research has been conducted on the cosmic and astrophysical consequences of non-minimal connections between matter and geometry \cite{ref37,ref38,ref39,ref40,ref41}. Harko and Lobo \cite{ref42}, recently introduced a more advanced version of matter-curvature coupling theories known as $f(R, L_{m})$ gravity theory \cite{R10,R4,R9}. This theory involves an arbitrary function of the matter Lagrangian density $L_m$ and the Ricci scalar $R$, making it the most extensive gravitational theory in Riemann space. In $f(R, L_{m})$ gravity theory, test particles do not follow geodesic motion and experience an additional force perpendicular to their four velocity vectors. The $f(R, L_{m})$ gravity models allow for a clear violation of the equivalence principle, which is closely limited by tests within the solar system \cite{ref43,ref44}. Wang and Liao have recently investigated energy conditions within the framework of $f(R, L_{m})$ gravity \cite{ref45}. Gonclaves and Moraes examined the field of cosmology by studying the $f(R, L_{m})$ gravity theory that involves coupling with non-minimal matter geometry \cite{ref46}. Various investigators have found that a significant amount of comoving entropy is produced during cosmological evolution as a result of the connection between curvature and matter in the universe \cite{ref47,ref48}. They have observed that $f(R, L_m)$ gravity could provide a better framework for comprehending the universe and its gravitational dynamics compared to GR. A few more interested papers can be consulted in this context under $f(Q, L_{m})$ gravity and $f(Q, T)$ gravity \cite{R6,R7,R3,R8}. 

Based on the above-mentioned works we are motivated to study cosmological scenario under the framework of $f(R,\mathcal{L}_{m})$ gravity with quadratic equation of state (EOS) as this quadratic case will allow us to understand the cosmic evolution in a wider range and effect of the EOS parameter on the evolutionary processes. Therefore, our investigation is structured as follows: in Sec. \ref{sec2}, we provide the metric and field equations under $f(R,\mathcal{L}_{m})$ gravity. In Sec. \ref{sec3}, we employ the observational data and relevant methodology to explore various features of our presented theoretical model to validate it with the established attributes. A verification scheme has been adopted in  Sec. \ref{sec4}  for achieving various cosmological features whereas in Sec. \ref{sec5} we discuss about energy conditions. Finally, we conclude our work in Sec. \ref{sec6}.

\section{The metric and field equations in  $f(R,\mathcal{L}_{m})$ gravity} \label{sec2}

The Einstein-Hilbert action for $f(R,\mathcal{L}_{m})$ gravity is given by \cite{ref42}
\begin{equation}
S = \int {f(R,\mathcal{L}_{m}) \sqrt{-g} d^{4}x}, \label{eq1}
\end{equation}
where $f(R,\mathcal{L}_{m})$ is an arbitrary function of Ricci scalar $R$ and matter Lagrangian $\mathcal{L}_{m}$.

In terms of Ricci-tensor $R_{ij}$ and metric-tensor $g_{ij}$, the Ricci scalar is defined as 
\begin{equation}
R = g^{ij} R_{ij}, \label{eq2}
\end{equation}
where Ricci-tensor is written as
\begin{equation}
R_{ij} = \partial_{k} \varGamma^{k}_{i \, j} - \partial_{j} \varGamma^{k}_{k \, i}+\varGamma^{\lambda}_{i \, j} \, \varGamma^{k}_{\lambda \, k} - \varGamma^{k}_{j \, \lambda} \, \varGamma^{\lambda}_{k \, i}. \label{eq3}
\end{equation}

The Levi-Civita connection $\varGamma^{\alpha}_{\beta \, \gamma}$ is expressed as
\begin{equation}
\varGamma^{\alpha}_{\beta \, \gamma} = \frac{1}{2} g^{\alpha \lambda} \bigg(\frac{\partial g_{\gamma \lambda}}{\partial x^\beta} +\frac{\partial g_{\lambda \beta}}{\partial x^\gamma}-\frac{\partial g_{\beta \gamma}}{\partial x^\lambda} \bigg). \label{eq4}
\end{equation}

On variation of action (\ref{eq1}) over metric tensor $g_{ij}$, the field equations for $f(R,\mathcal{L}_{m})$ gravity given as \cite{ref36}:
\begin{equation}
\frac{\partial f}{\partial R} R_{ij}+\left(g_{ij} \square-\nabla_{i} \nabla_{j}\right) \frac{\partial f}{\partial R}-\frac{1}{2}\bigg(f-\frac{\partial f}{\partial \mathcal{L}_{m}} \mathcal{L}_{m}\bigg) g_{ij} = \frac{1}{2} \bigg(\frac{\partial f}{\partial \mathcal{L}_{m}}\bigg) T_{ij}, \label{eq5}
\end{equation}
where $T_{ij}$ is the energy-momentum tensor of perfect fluid and is expressed as
\begin{equation}
T_{ij} = \frac{-2}{\sqrt{- g}} \frac{\delta (\sqrt{-g} \mathcal{L}_{m})}{\delta g^{ij}},\label{eq6}
\end{equation}

By contracting the field equation  (\ref{eq5}), we establish the relationship connecting Ricci scalar ($R$), the trace of energy-momentum tensor ($T$) and matter Lagrangian density ($\mathcal{L}_{m}$) as
\begin{equation}
\frac{\partial f}{\partial R} R+3 \, \square \frac{\partial f}{\partial R}-2\bigg(f-\frac{\partial f}{\partial \mathcal{L}_{m}} \mathcal{L}_{m}\bigg) = \frac{1}{2} \bigg(\frac{\partial f}{\partial \mathcal{L}_{m}}\bigg) T, \label{eq7}
\end{equation}
where $\square u = \frac{1}{\sqrt{-g}} \partial_{i} (\sqrt{-g} g^{ij} \partial_{j} u) $ for an arbitrary function $u$.

Taking covariant derivative of Eq. (\ref{eq5}), we get
\begin{equation}
\nabla^{i} T_{ij} = 2 \nabla^{i} log\bigg(\frac{\partial f}{\partial \mathcal{L}_{m}}\bigg) \frac{\partial \mathcal{L}_{m}}{\partial g^{ij}}, \label{eq8}
\end{equation}

In the present study, we consider the flat FLRW space-time metric representing a homogeneous and isotropic universe: 
\begin{equation}
ds^2= a^2(t) \left( dx^2+dy^2+ dz^2 \right)-dt^2, \label{eq9}
\end{equation}
where $a(t)$ denote scalar factor describing the expansion of universe with time $t$. 

The Christoffel symbols exhibit non-zero components as follows:
\begin{equation}
\varGamma^{0}_{i \, j} = -\frac{1}{2} g^{00}\frac{\partial g_{ij}}{\partial x^0}, \, \, \varGamma^{k}_{0 \, j} =\varGamma^{k}_{j \, 0}= \frac{1}{2} g^{k \lambda}\frac{\partial g_{j \lambda}}{\partial x^0}, \label{eq10}
\end{equation}
where $i, j, k = 1, 2, 3$.

Using Eq. (\ref{eq3}), the non-zero components of Ricci tensor are obtained as:
\begin{equation}
R_{00} = 3 \frac{\ddot{a}}{a}, \,\,\,  R_{11} = R_{22} = R_{33} = \frac{\ddot{a}}{a}+2 \frac{\dot{a}^2}{a^2}, \label{eq11}
\end{equation}

Thus, the Ricci scalar for the line element (\ref{eq9}) is 
\begin{equation}
R = 6 \left(\frac{\ddot{a}}{a} + \frac{\dot{a}^2}{a^2}\right)= 12 H^2+6\dot{H}, \label{eq12}
\end{equation}
where $H = \frac{\dot{a}}{a}$ is the Hubble parameter.

Assuming that the universe's matter distribution is characterized by the energy-momentum tensor of perfect fluid. Therefore, for line element (\ref{eq9}), the energy-momentum tensor of perfect fluid is given as
\begin{equation}
T_{ij} = (\rho + p)u_{i}u_{j} + p g_{ij}, \label{eq13}
\end{equation}
where $\rho$ and $p$ denote the energy density and thermodynamic pressure of matter, respectively. Also, $g^{ij}$ is the metric tensor and $u^{i} = (1,0,0,0)$ is the component of four velocity vector which satisfying $u_{i}u^{i} = -1$.

In the background of $f(R,\mathcal{L}_{m})$ theory, the modified Friedmann equations depicting the dynamics of universe can be read as:
\begin{equation}
R_{00} \frac{\partial f}{\partial R}+\frac{1}{2} \left(\mathcal{L}_{m}\frac{\partial f}{\partial \mathcal{L}_{m}}-f\right)+3 H \frac{\partial \dot{f}}{\partial R} = \frac{1}{2} \frac{\partial f}{\partial \mathcal{L}_{m}} T_{00} \label{eq14}
\end{equation}
and
\begin{equation}
R_{11} \frac{\partial f}{\partial R}+\frac{1}{2} \left(\mathcal{L}_{m}\frac{\partial f}{\partial \mathcal{L}_{m}}-f\right)-\frac{\partial \ddot{f}}{\partial R}-3 H \frac{\partial \dot{f}}{\partial R} = \frac{1}{2} \frac{\partial f}{\partial \mathcal{L}_{m}} T_{11}. \label{eq15}
\end{equation}

Now, we consider the $f(R,\mathcal{L}_{m})$ function as \cite{ref49}
\begin{equation}
f(R,\mathcal{L}_{m}) = \frac{R}{2}+\mathcal{L}^{n}_{m}, \label{eq16}
\end{equation}
where $n$ is arbitrary constant.

In the specific scenario of this $f(R,\mathcal{L}_{m})$ gravity model with $\mathcal{L}_{m} = \rho$ \cite{ref50}, the Friedmann equations (\ref{eq14})--(\ref{eq15}) are transformed as 
\begin{equation}
3 H^2 = (2n-1) \rho^{n}, \label{eq17}
\end{equation}

\begin{equation}
2\dot{H}+3 H^2 = (n-1) \rho^{n}- n \rho^{n-1}p. \label{eq18}
\end{equation}

The standard Friedmann equations of general relativity (GR) can be retrieved from above Eqs. (\ref{eq17})--(\ref{eq18}) for $n=1$.

For the proposed model, $\omega = \frac{p}{\rho}$ is the equation of state (EOS) parameter. From Eqs. (\ref{eq17})--(\ref{eq18}), EoS parameter $\omega$ is recasts as:
\begin{equation}
\omega=\frac{2 (2n-1)\dot{H}+3nH^2}{-3nH^2}. \label{eq19}
\end{equation}

Utilizing the relation $1+z = \frac{\dot{a}}{a}$, we get $\frac{\dot{H}}{H}=-(1+z) H'$ and Eq. (\ref{eq19}) is transformed into 
\begin{equation}
\omega=\frac{-2(2n-1)(1+z) H H'+3nH^2}{-3nH^2}.
\end{equation}

For the cosmological analysis, we consider the quadratic form of EOS parameter as $\omega(z)=-1 + \alpha (1 + z) + \beta (1 + z)^2$, where $\alpha$ and $\beta$ are constants. As a general case, this redshift-quadratic EOS parameter allows for a more dynamic description of dark energy's evolution with redshift ($z$) compared to a constant EOS. This approach, where $\alpha$ and $\beta$ are constants, helps scientists to understand the universe's expansion history and potential deviations from the standard $\Lambda$CDM model. For $\beta = 0$, it turns into a linear EOS parameterization model, while for $\alpha = 0$ and $\beta = 0$ it approaches to a standard $\Lambda$CDM model ($\omega=-1$).

Thus, in terms of redshift, the Hubble parameter can be expressed as~\cite{WA}:
\begin{equation}
 H = H_{0} \, e^{\frac{3 n z \left(\alpha +\beta +\frac{\beta  z}{2}\right)}{2 (2 n-1)}},
\end{equation}
where $H_0$ represents the present value of Hubble parameter.

The deceleration parameter of the model is read as
\begin{equation}
q=-1 + \frac{d}{dt}\left(\frac{1}{H}\right).
\end{equation}

\section{Observational data and relevant methodology for cosmic evolutionary scenario} \label{sec3}

Our primary goal in this section is to determine the free parameter of the derived universe model using the $\chi^2$-minimization method with observational data sets. To find the values of $\alpha$, $\beta$, $n$, and $H_{0}$ in the model, we utilize the data sets listed below: \\

\noindent \textbf{(i) CC}: From the cosmic chronometric approach, we have collected $31~H(z)$ observational data points in the interval $0.07\leq z\leq 1.965$ \cite{ref51,ref52,ref53}.\\
 
\noindent\textbf{(ii) Pantheon}: In this work, we use the recent Pantheon compilation of SN Ia data in the redshift range $0.001 < z < 2.26$ \cite{ref54}. The Pantheon data is compiled from 18 different surveys, but with the SNe Ia light curves now uniformly fit to SALT2 model light curves \cite{ref55}.\\ 
 
\noindent \textbf{(iii) BAO data}: We also considered the six BAO data points \cite{ref56,ref57,ref58}. For the BAO sample, the predictions from a sample of Galaxy Surveys like SDSS DR7 and 6dF, and WiggleZ have been utilized\cite{ref56,ref57,ref58}. The angular diameter distance for the sample is defined as $D_{A}=\frac{D_{L}}{(1+z)2}$, where $D_{L}$ indicates the proper angular diameter distance \cite{ref59}, and the dilation scale is described by $D_{V}(z)=\left[ D^{2}_{L}(z)*(1+z)^2*\frac{c \,  z}{H(z)} \right]^{1/3}$. 

For limiting the parameters of the model,  the chi-square estimator for the BAO sample is described in the following form \cite{ref59,ref60,ref61,ref62}.
\begin{equation}
\chi^2_{BAO} = X^{T} C^{-1} X,
\end{equation}
where 
\begin{align*}
X &= \begin{pmatrix}
\frac{d_A(z_{*})}{D_V(0.106)}-30.95\\           
\frac{d_A(z_{*})}{D_V(0.20)}-17.55\\
\frac{d_A(z_{*})}{D_V(0.35)}-10.11\\
\frac{d_A(z_{*})}{D_V(0.44)}-8.44\\
\frac{d_A(z_{*})}{D_V(0.60)}-6.69\\
\frac{d_A(z_{*})}{D_V(0.73)}-5.45
\end{pmatrix}
\end{align*}

%%%%%%%%%%%%%%%%%%%%%%%%%%%%%%%%%%%%%%%%%%%%%%%%%%%%%%%%%%%%%%%%%%%%%%%%%%%% Table 1 %%%%%%%%%%%%%%%%%%%%%%%%%%%%%%%%%%%%%%%%%%%%%%%%%%%%%%%%%%%%%%%%%%%%%%%%%%%%%%%%%%
\begin{table}
\caption{ The BAO dataset consisting values of $\varUpsilon(z)$ for different points of $z_{BAO}$ with $\varUpsilon(z)= d_A(z_{*})/D_V(z_{BAO})$ and $z_{*}\approx 1091$.}
\begin{center}
\begin{tabular}{|c|c|c|c|c|c|c|}
\hline 
%\multicolumn{7}{|c|}{The values of $\varUpsilon(z)$ for different points of $z_{BAO}$}\\
\hline
\tiny $z_{BAO}$ & \tiny $0.106$ & \tiny $0.2$ & \tiny $0.35$  & \tiny $0.44$ & \tiny $0.6$ & \tiny $ 0.73$  \\
\hline
\tiny   $\varUpsilon(z)$ & \tiny $30.95 \pm 1.46$ & \tiny $17.55 \pm 0.60$  & \tiny $10.11 \pm 0.37$ & \tiny  $8.44 \pm 0.67$ & \tiny  $6.69 \pm 0.33$ & \tiny $ 5.45 \pm 0.31$\\
\hline
\end{tabular}
\end{center}
\end{table}
%%%%%%%%%%%%%%%%%%%%%%%%%%%%%%%%%%%%%%%%%%%%%%%%%%%%%%%%%%%%%%%%%%%%%%%%%%%%%%%%%%%%%%%%%%%%%%%%%%%%%%%%%%%%%%%%%%%%%%%%%%%%%%%%%%%%%%%%%%%%%%%%%%%%%%%%%%%%%%%%%%%%%%%

The model parameters $\alpha$, $\beta$, $n$, and $H_{0}$ of this model can be assessed using observational data and statistical methods. $E_{obs}$ represents the value of any observational data, while $E_{th}$ represents the values that have been theoretically estimated in the model of the Universe. By using statistical methods to compare the two values $E_{obs}$ and $E_{th}$, the model's parameters can be determined. To obtain the most accurate estimates, we utilized the $\chi^2$ estimator. If the standard error in the observed values is denoted by $\sigma$, the formula for the $\chi^2$ estimator is expressed as
\begin{equation}
\label{chi1}
\chi^{2} = \sum_{i=1}^{N}\left[\frac{E_{th}(z_{i})- E_{obs}(z_{i})}{\sigma_{i}}\right]^{2},
\end{equation}
where $E_{th}(z_{i})$ represents the theoretical values of the parameter, while $E_{obs}(z_{i})$ represents the observed values. $\sigma_{i}$ is the error and N is the number of data points.

The combined $\chi^{2}$ estimator is read as
\begin{equation}
\label{chi2}
\chi^{2}_{combined} = \chi^{2}_{CC} + \chi^{2}_{Pantheon} + \chi^{2}_{BAO}.
\end{equation}

%%%%%%%%%%%%%%%%%%%%%%%%%%%%%%%%%%%%%%%%% Figure 1%%%%%%%%%%%%%%%%%%%%%%%%%%%%%%%%%%%%%%%%
\begin{figure}
\centering
\includegraphics[scale=0.4]{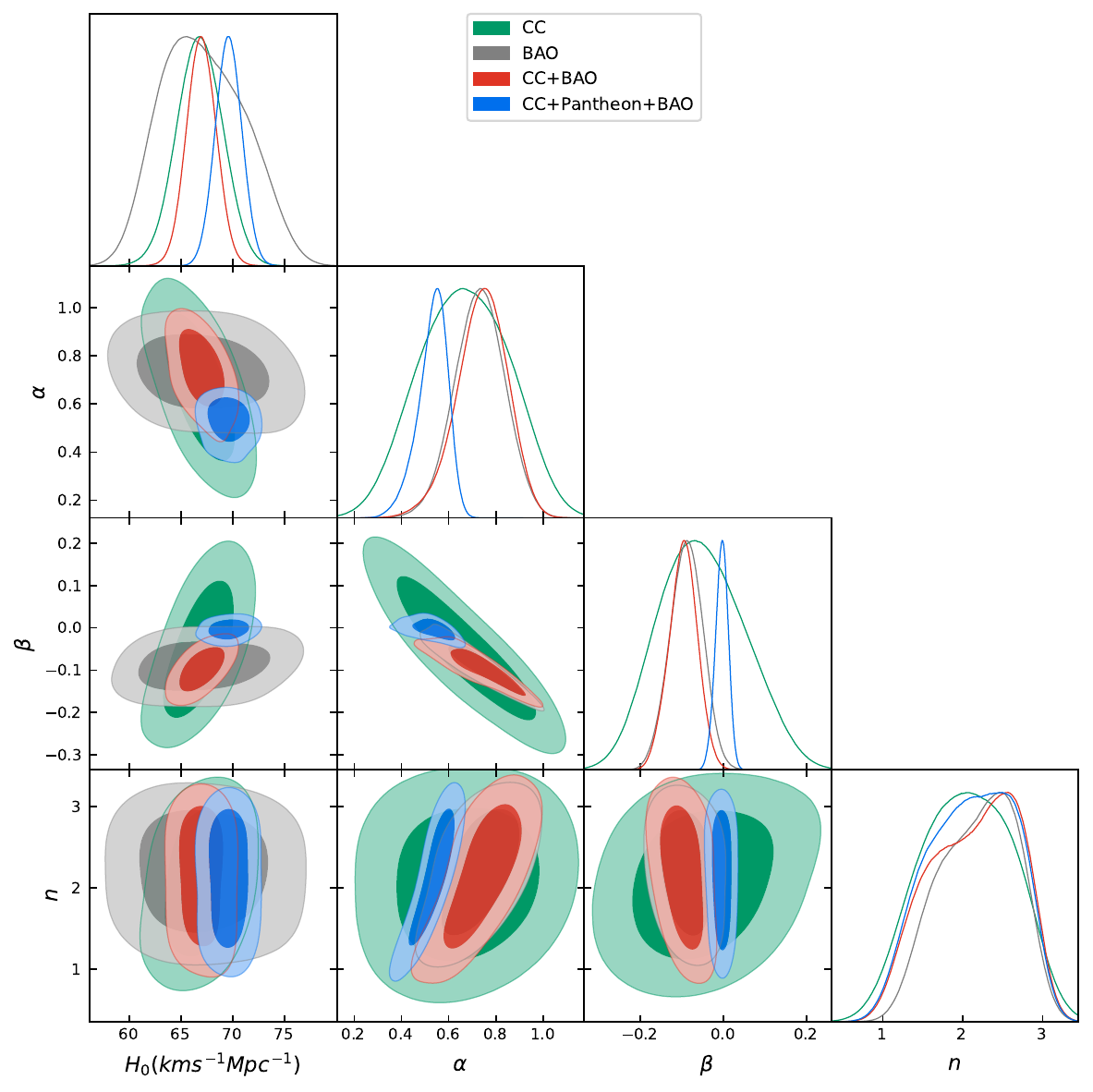}
\caption{One-dimensional marginalized distribution, and $2D$ contours with $1\sigma$ and $2\sigma$ confidence levels for the derived model of the Universe with CC, Pantheon, BAO, and joint datasets.}
\end{figure}
%%%%%%%%%%%%%%%%%%%%%%%%%%%%%%%%%%%%%%%%%%%%%%%%%%%%%%%%%%%%%%%%%%%%%%%%%%%%%%%%%%%%%%%%%%

%%%%%%%%%%%%%%%%%%%%%%%%%%%%%%%%%%%%%%%%% Figure 2 %%%%%%%%%%%%%%%%%%%%%%%%%%%%%%%%%%%%%%%
\begin{figure}
	\centering
	(a)\includegraphics[width=7cm,height=5cm,angle=0]{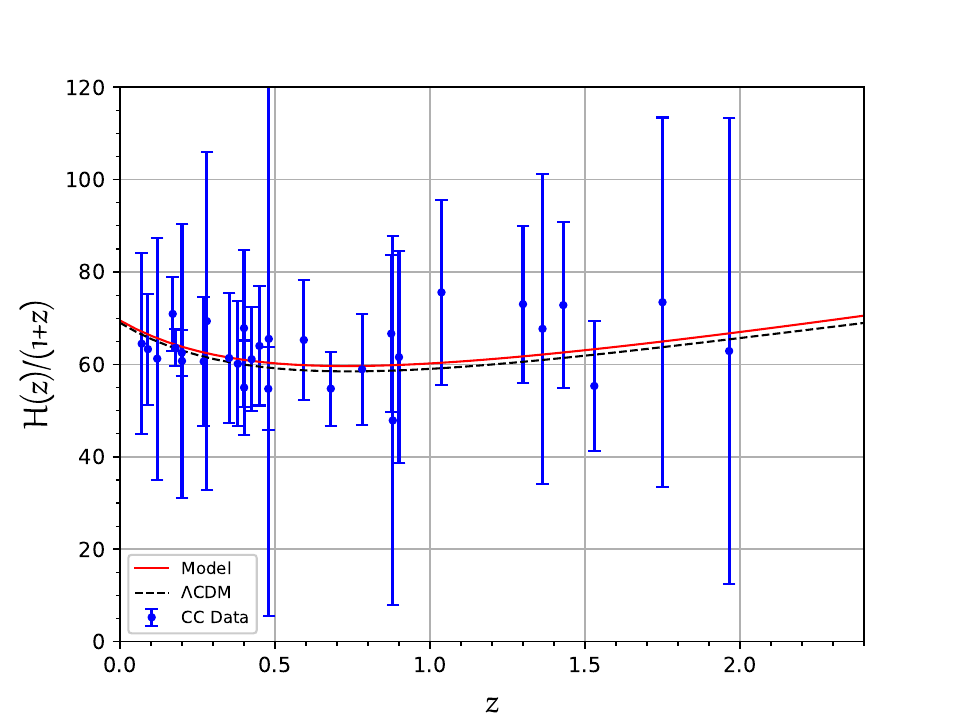}
	(b)\includegraphics[width=7cm,height=5cm,angle=0]{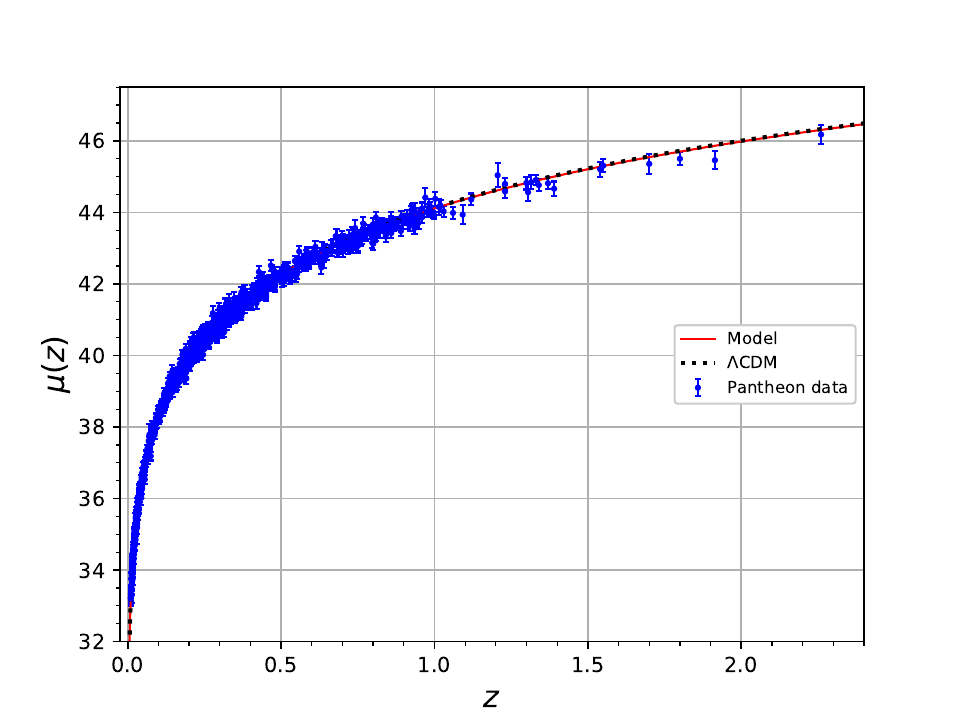}
	\caption{(a) Profile of Hubble rate for $f(R,\mathcal{L}_{m})$ model (red line), Hubble rate plot for $\Lambda$CDM model (black dashed), and CC dataset with error-bars (blue dots), (b) Distance modulus plot for $f(R,\mathcal{L}_{m})$ model (red line) and $\Lambda$CDM model (black dashed) vs Pantheon dataset with error-bars (blue dots). } 
\end{figure}
%%%%%%%%%%%%%%%%%%%%%%%%%%%%%%%%%%%%%%%%%%%%%%%%%%%%%%%%%%%%%%%%%%%%%%%%%%%%%%%%%%%%%%%%%%%%%%%%%%%%%%%%%%%%%%%%%%%%%%%%%%%%%%%%%%%%%%%%%%%%%%%%%%%%%%%%%%%%%%%%%%
\section{Verification of the model's cosmological features.} \label{sec4}

%%%%%%%%%%%%%%%%%%%%%%%%%%%%%%%%%%%%%%%%%%%%%%%%%%%%%%%%%%%%%%%%%%%%%%%%%%%%%%%%%%%%%%%%%%%%%%%%%%%%%%%%%%%%%%%%%%%%%%%%%%%%%%%%%%%%%%%%%%%%%%%%%%%%%%%%%%%%%%%%%%%%%%%
\begin{figure}
	\centering
	(a)\includegraphics[width=7cm,height=5cm,angle=0]{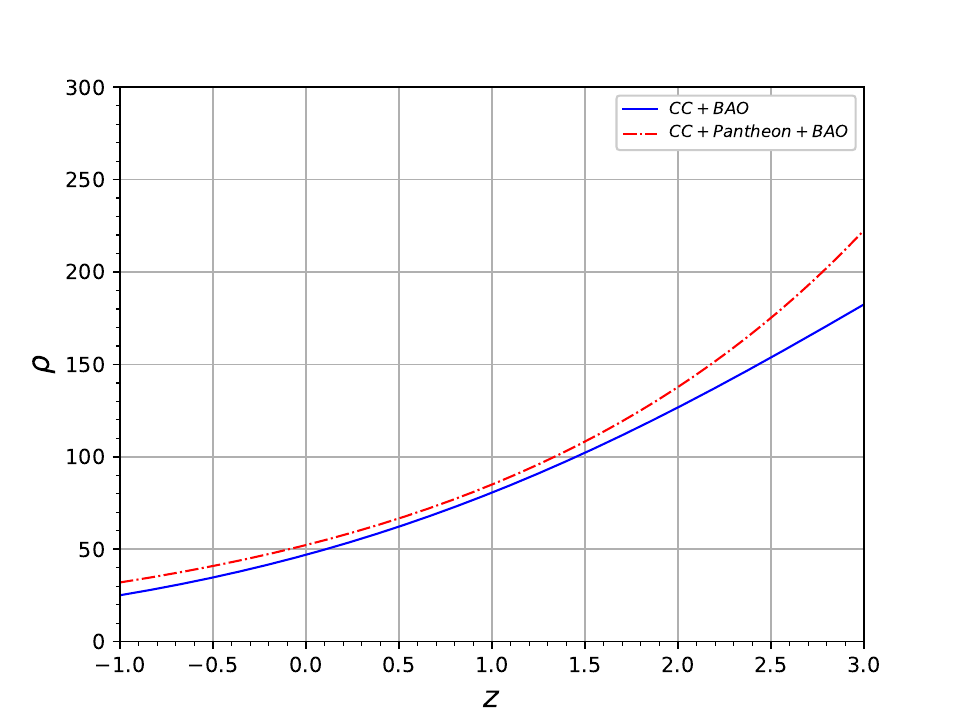}
	(b) \includegraphics[width=7cm,height=5cm,angle=0]{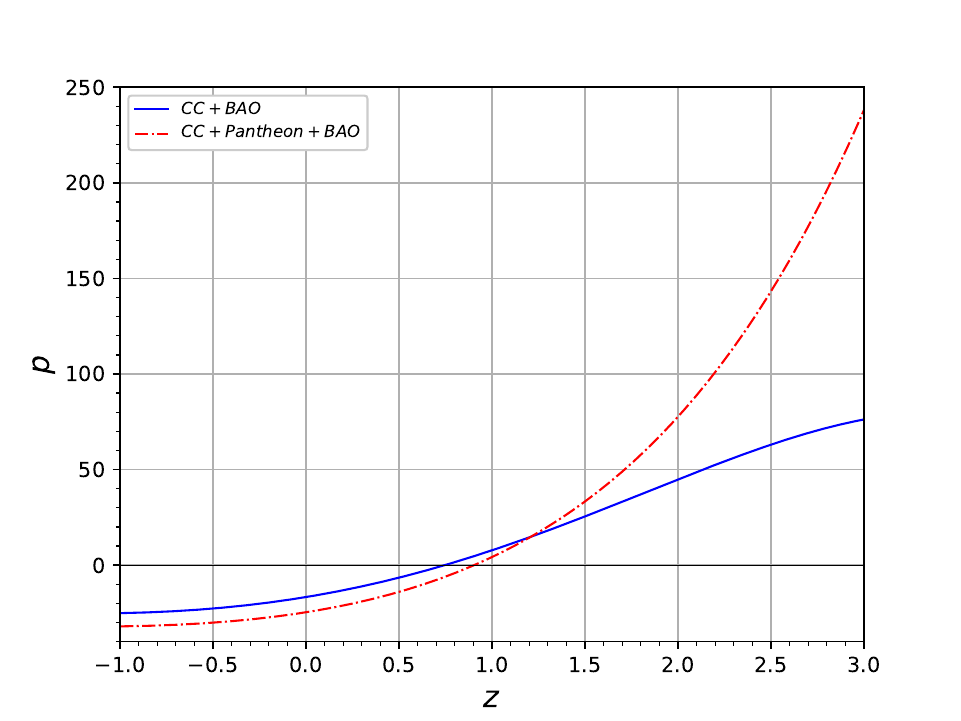}
	\caption{(a) plot of energy density, (b) plot of cosmic pressure.}
\end{figure}
%%%%%%%%%%%%%%%%%%%%%%%%%%%%%%%%%%%%%%%%%%%%%%%%%%%%%%%%%%%%%%%%%%%%%%%%%%%%%%%%%%%%%%%%%%%%%%%%%%%%%%%%%%%%%%%%%%%%%%%%%%%%%%%%%%%%%%%%%%%%%%%%%%%%%%%%%%%%%%%%%%%%%%%

\begin{eqnarray}
\rho &=&  3^{1/n} \left(\frac{H_0^2 e^{\frac{3 n z \left(\alpha +\beta +\frac{\beta  z}{2}\right)}{2 n-1}}}{2 n-1}\right)^{1/n},
\end{eqnarray}

\begin{eqnarray}
p &=& \left(\alpha  (z+1)+\beta  (z+1)^2-1\right) \left(\frac{ 3 H_0^2 e^{\frac{3 n z \left(\alpha +\beta +\frac{\beta  z}{2}\right)}{2 n-1}}}{2 n-1}\right)^{1/n}.    \end{eqnarray}

For the derived model in $f(R,\mathcal{L}_{m})$ gravity, the energy density is positive through the entire evolution of the universe. Trajectory of energy density can be seen in Fig. 3 (a). The cosmic pressure for the proposed model is initially positive that indicates a matter dominated era of the universe. But later on, the pressure becomes negative in later era of universe evolution which represent the dark energy dominance. The behavior of cosmic pressure for the proposed model is plotted in Fig. 3(b). 
%%%%%%%%%%%%%%%%%%%%%%%%%%%%%%%%%%%%%%%% Table 2 %%%%%%%%%%%%%%%%%%%%%%%%%%%%%%%%%
\begin{table}
	\caption{The assessed values of parameters for various observational sets where I, II, III and IV stand for CC, BAO, CC+BAO and CC+Pantheon+BAO respectively.}
	\begin{center}
		\begin{tabular}{|c|c|c|c|c|}
		\hline\hline 
		\tiny Parameters & \tiny $H_{0}$ $(km s^{-1} Mpc^{-1})$ & \tiny $\alpha$	& \tiny $\beta$ & \tiny $n$  \\
		\hline
		\tiny I & \tiny $66.808949 \pm 2.053332$ & \tiny $0.667865 \pm 0.206599$ & \tiny $-0.051849 \pm 0.109246$ & \tiny $2.058320 \pm 0.637458$   \\ 
		\hline
		\tiny II & \tiny $67.103341 \pm 4.249948$ & \tiny $ 0.731479 \pm 0.108153$ & \tiny $-0.089242 \pm 0.038624$ & \tiny $ 2.208137 \pm 0.558474$   \\
		\hline
		\tiny III & \tiny $66.963076 \pm 1.346302$ & \tiny $0.741964 \pm 0.100818$ & \tiny $-0.096889 \pm 0.034385$ & \tiny $2.157851 \pm 0.647475$  \\
		\hline
		\tiny IV & \tiny $69.594080 \pm 1.235029$ & \tiny $0.532354 \pm 0.061815$ & \tiny $-0.003720 \pm 0.014789$ & \tiny $2.124560 \pm 0.626690$  \\
		\hline\hline
		\end{tabular}
	\end{center}
\end{table}

%%%%%%%%%%%%%%%%%%%%%%%%%%%%%%%%%%%%%%%%%%%%%%%%%%%%%%%%%%%%%%%%%%%%%%%%%%%%%%%%%%%%%%%%%%%
%%%%%%%%%%%%%%%%%%%%%%%%%%%%%%%%%%%%%%%%%%%%%%%%%%%%%%%%%%%%%%%%%%%%%%%%%%%%%%%%%%%%%%%%%%%%%%%%%%%%%%%%%%%%%%%%%%%%%%%%%%%%%%%%%%%%%%%%%%%%%%%%%%%%%%%%%%%%%%%%%%%%%%%
\begin{figure}
	\centering
	\includegraphics[scale=0.5]{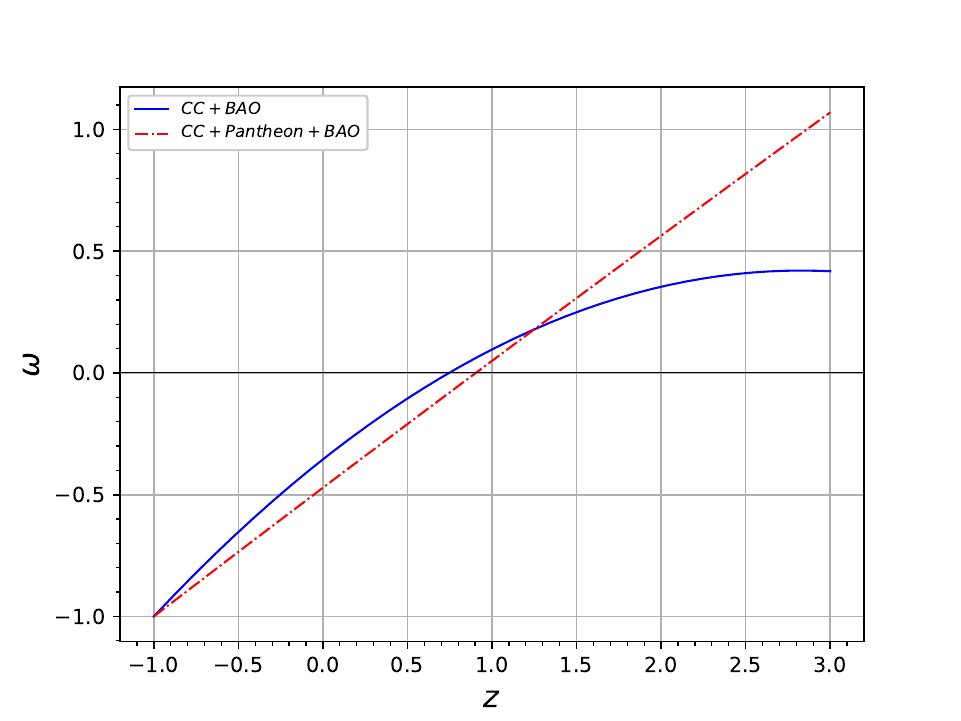}
	\caption{ Graphical plot of EOS parameter.}
\end{figure}
%%%%%%%%%%%%%%%%%%%%%%%%%%%%%%%%%%%%%%%%%%%%%%%%%%%%%%%%%%%%%%%%%%%%%%%%%%%%%%%%%%%%%%%%%%%%%%%%%%%%%%%%%%%%%%%%%%%%%%%%%%%%%%%%%%%%%%%%%%%%%%%%%%%%%%%%%%%%%%%%%%%%%%%

For $\omega > 0$, the EOS parameter describes the Big Bang era of the universe, and $\omega = 0$ represents a matter-dominated scenario of the cosmos. In this case, dark matter is a non-relativistic matter for which $\omega= 0$, while for a relativistic matter like radiation $\omega= 1/3$. The model lies in the quintessence era for $-1< \omega \leq 0$. In the case of $\omega = -1$, the EOS parameter approaches to the $\Lambda$CDM model. For $\omega < -1$, a phantom era is observed. We have plotted the trajectories of the EOS parameter in Fig. 4 for the derived model. This figure shows that the model begins with the initial Big Bang era crosses the quintessence era and approaches to $\Lambda$CDM in late-time phase.

\subsection{Deceleration parameter}

One of the key factors that describes the universe’s evolutionary features is the deceleration parameter (DP) $q$, which plays a crucial role in determining the universe's phase transition. From Hubble parameter, DP can be derived as $q= -1 + \frac{(1+z)}{H(z)} \frac{d H(z)}{dz} $. Therefore, the DP $q$ for the suggested model can be recasts as:
\begin{equation}
q =  \frac{n \left(3 \alpha  (z+1)+3 \beta  (z+1)^2-4\right)+2}{4 n-2}.
\end{equation}

%%%%%%%%%%%%%%%%%%%%%%%%%%%%%% Figure 3 %%%%%%%%%%%%%%%%%%%%%%%%%%%%%%%%%%%%%%%%%%%%%%%%
\begin{figure}
	\centering
	\includegraphics[scale=0.5]{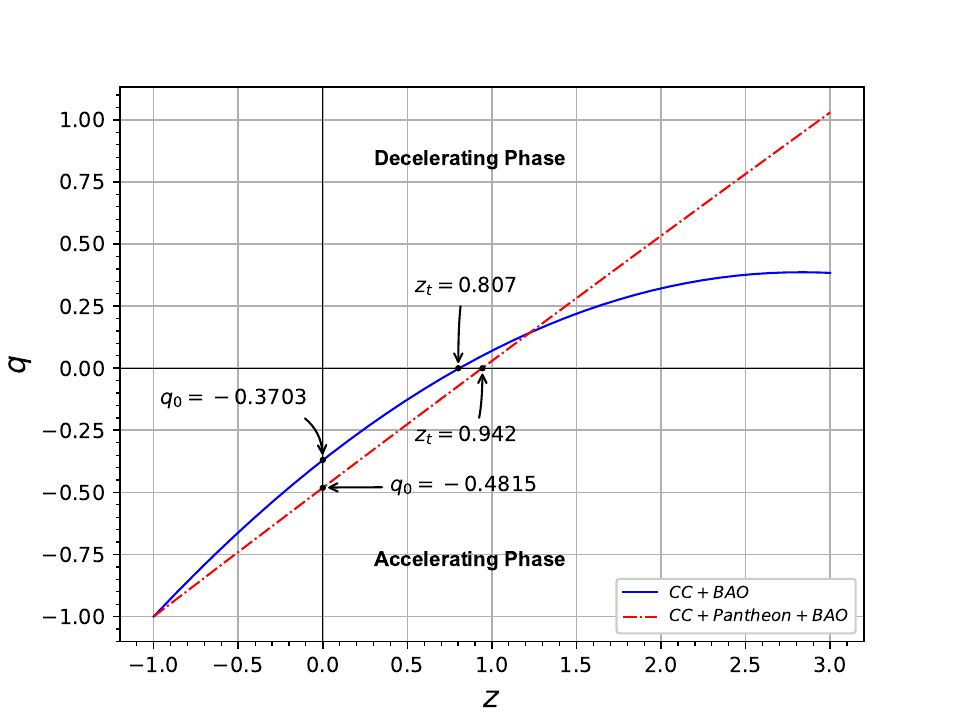}
	\caption{ Graphical plot of Deceleration parameter.}
\end{figure}
%%%%%%%%%%%%%%%%%%%%%%%%%%%%%%%%%%%%%%%%%%%%%%%%%%%%%%%%%%%%%%%%%%%%%%%%%%%%%%%%%%%%%%%%%
%%%%%%%%%%%%%%%%%%%%%%%%%%%%%%%%%%%%%%% Table 2 %%%%%%%%%%%%%%%%%%%%%%%%%%%%%%%%%%%%%%%
\begin{table}
	\caption{ Table of numerical findings derived from the suggested $f(R,\mathcal{L}_{m})$ model where I, II, III, IV and V stand for CC, BAO, Pantheon, CC+BAO and CC+Pantheon+BAO respectively. }
	\begin{center}
		\begin{tabular}{|c|c|c|c|c|c|}
			\hline\hline 
			\tiny Parameters & \tiny I  & \tiny II	& \tiny III & \tiny IV & \tiny V   \\
			\hline
			\tiny $q_{0}$ & \tiny $-0.3897^{+0.2478}_{-0.2629}$ & \tiny $-0.3773^{+0.0996}_{-0.0895}$  & \tiny $-0.5866^{+0.1146}_{-0.1171}$ & \tiny $-0.3703^{+0.0824}_{-0.0581}$ & \tiny $-0.4815^{+0.0362}_{-0.0096}$    \\ 
			\hline
			\tiny $z_{t}$ & \tiny $0.749^{+1.122}_{-0.594}$ & \tiny $0.810^{+1.032}_{-0.386}$  & \tiny $0.646^{+0.245}_{-0.191}$ & \tiny $0.807^{+0.899}_{-0.349}$ & \tiny $0.942^{+0.112}_{-0.164}$ \\
			\hline
			\tiny $j_{0}$ & \tiny $0.473^{+0.336}_{-0.113}$ & \tiny $0.444^{+0.109}_{-0.079}$  & \tiny $0.815^{+0.036}_{-0.029}$ & \tiny $0.439^{+0.094}_{-0.076}$ & \tiny $0.497^{+0.019}_{-0.018}$ \\
			\hline
			\tiny $\omega_{0} $ & \tiny $-0.384 \pm 0.316 $ & \tiny $-0.358 \pm 0.147$ & \tiny $-0.636 \pm 0.150$ & \tiny $-0.355 \pm 0.135 $ & \tiny $-0.471\pm 0.077$ \\
			\hline
		\end{tabular}
	\end{center}
\end{table}
%%%%%%%%%%%%%%%%%%%%%%%%%%%%%%%%%%%%%%%%%%%%%%%%%%%%%%%%%%%%%%%%%%%%%%%%%%%%%%%%%%%%%%%%%%%%

The analysis of the deceleration parameter $q$ in conjunction with the Hubble parameter $H$ provides a comprehensive understanding of the most of expansion characteristics of the universe. Both the age of the universe and its phase transitions can be effectively characterized using $H$ and $q$.

Capozziello et al. \cite{ref63}, found redshift transitions at $z_{t}={0.646}^{+0.020}_{-0.158}$, $z_{t}={0.659}^{+0.371}_{-0.124}$, $z_{t}={0.860}^{+0.013}_{-0.146}$, and $z_{t}={1.183}^{+0.002}_{-0.032}$ using SNIa, OHD, BAO, and joint datasets. Xu and Liu \cite{ref64}, recently found that the current value of DP is $q_{0}=-0.31 \pm 0.72$ for OHD data, while Giostri et al. \cite{ref59}, calculated $q_{0}=-0.31 \pm 0.11$ and $q_{0}=-0.31 \pm 0.20$ for BAO and CMB data. Crevecoeur \cite{ref65} estimated $q_{0}$ to be approximately $-0.25$ using Planck's data. Giostri et al. \cite{ref59}, found the current value of DP to be $q_0=-0.53^{+0.17}_{-0.13}$ through the constraint of SNIa and BAO/CMB datasets. Similarly, Santos et al. \cite{ref66}, also determine $q_0=-0.54^{+0.05}_{-0.07}$ by analyzing information from $H(z)$, SNIa, and BAO/CMB. 

In our present work, the model depicts a flipping nature of the universe from a deceleration era to a current expansion scenario that indicates a dark energy (DE) dominance in the present while a matter dominance in the past. For the joint dataset of CC, BAO, and Pantheon, the proposed model shows a transition behavior with the present value of DP as $q_0 = -0.4815^{+0.0362}_{-0.0096}$ and signature flipping occurring at $z_{t} = 0.942^{+0.112}_{-0.164}$ as shown in Figure 5. The present analysis also predicts that this trend will continue as $z$ approaches $-1$ and $q$ approaches $-1$. The current value of the deceleration parameter in the proposed model for Pantheon data is calculated as $-0.5866^{+0.1146}_{-0.1171}$ aligning well with the most recent observational findings. The values of $q_{0}$ and $z_{t}$ calculated for various data sets are listed in Table 2. The results obtained from the model align well with the latest experimental findings \cite{ref51,ref64,ref66,ref67,ref68,ref69,ref70,ref71,ref72}.

\subsection{Statefinders diagnostic}

Hubble parameter $H$ and deceleration parameter $q$ together provide a complete explanation of the universe's evolutionary dynamics. Both parameters have the scale factor $a$, as well as its derivatives of first and second orders. 
This reliance causes all the suggested models to converge towards a common value of $q$ along with other key parameters, which diminishes the precision attributes of the theoretical models proposed. Important predictions regarding the precision of results from theoretical models are overlooked in the process. Two novel parameters $r$ and $s$ named as statefinder are introduced to distinguish the accuracy levels of various cosmological models with dark energy.\\

The statefinder pair assists in improving the precision of model predictions by pinpointing the evolutionary path in the $r-s$ plane. By considering various forms of dark energy, as mentioned in the literature \cite{ref73,ref74,ref75}, the distinction between the proposed cosmological model and the $\Lambda CDM$ model can be clearly distinguished on the $(r - ~s)$ plane \cite{ref73,ref76}. For the derived model, the parameters $r$ and $s$ in the standard forms can be recast as:
\begin{eqnarray}
r&=&2 q^{2}+q - \frac{\dot{q}}{H}\nonumber\\
&=& \bigg[n^2 (z+1)^2 \bigg(9 \alpha ^2-6 \beta +6 \alpha  \left(3 \beta  (z+1)-\frac{2}{z+1}\right)\nonumber\\
&+&9 \beta ^2 (z+1)^2+8\bigg)+n (z+1) \bigg(6 \alpha +3 \beta  (z+1)\nonumber\\
&-&\frac{8}{z+1}\bigg)+2\bigg] / \bigg[2 (1-2 n)^2\bigg],
\end{eqnarray}

\begin{eqnarray}
s&=&\frac{r-1}{3 (q-\frac{1}{2})}\nonumber\\
&=& \bigg[n (z+1) \bigg(3 \alpha ^2 n (z+1)+2 \alpha  \left(n \left(3 \beta  (z+1)^2-2\right)+1\right)\nonumber\\
&+& \beta  (z+1) \left(n \left(3 \beta  (z+1)^2-2\right)+1\right)\bigg)\bigg]\nonumber\\
& /& \bigg[3 (2 n-1) \left(n \left(\alpha  (z+1)+\beta  (z+1)^2-2\right)+1\right)\bigg].
\end{eqnarray}

Figure 6 illustrates the features of the suggested model in the $r-s$ plane using formulas for $r$ and $s$. The plot shows that the suggested model is situated within the Chaplygin gas region $(r > 1, s < 0)$ and converges to the $\Lambda$CDM point $(r = 1, s = 0)$ at the late passes of the time line. For joint dataset of CC, BAO, and Pantheon, the present values of parameters ($r, s$) are computed as (0.497, 0.171). Hence, the resulting model exhibits characteristics resembling $\Lambda$CDM in the current period, although not completely conforming to it.  

%%%%%%%%%%%%%%%%%%%%%%%%%%%%%%%%%%% Figure 4 %%%%%%%%%%%%%%%%%%%%%%%%%%%%%%%%%%%%%%%%%%%%
\begin{figure}
	\centering
	(a)\includegraphics[width=7.0cm,height=5.0cm,angle=0]{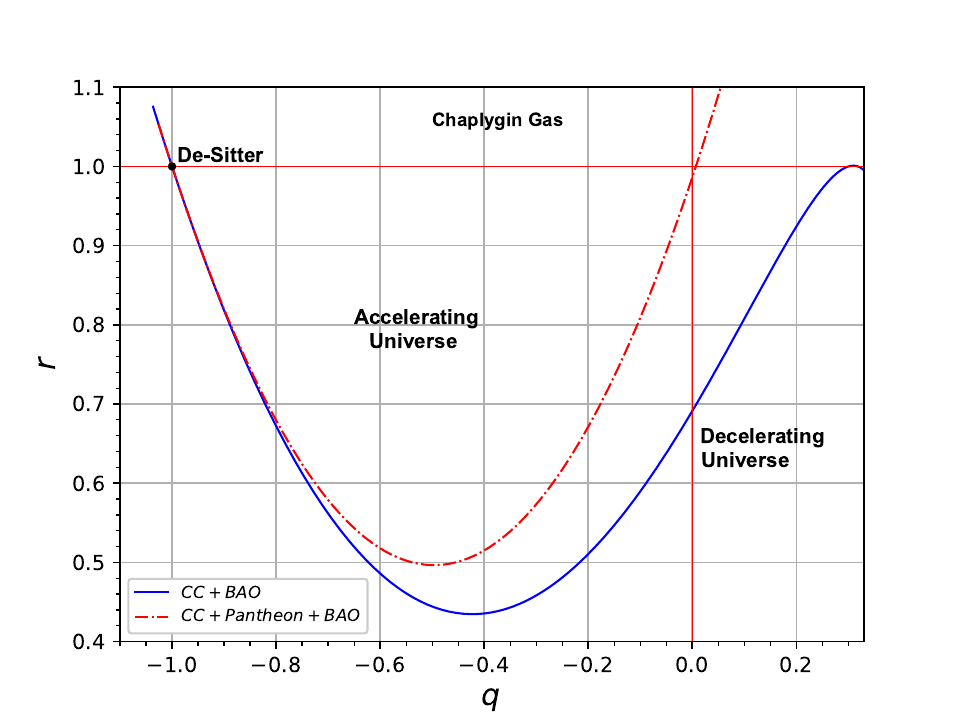}
	(b) \includegraphics[width=7.0cm,height=5.0cm,angle=0]{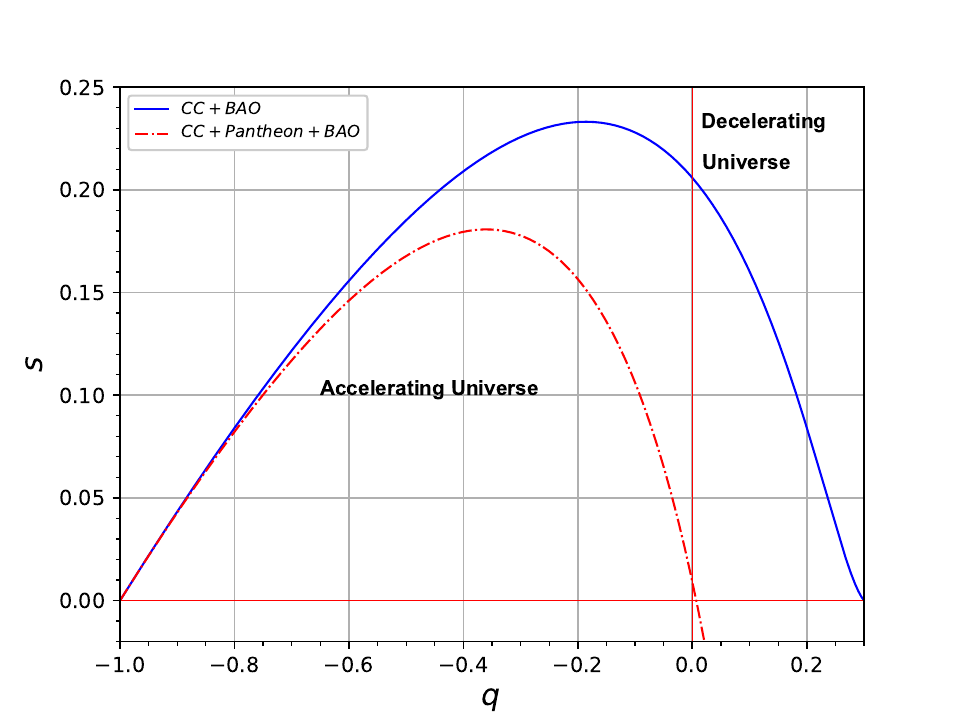}
	(c) \includegraphics[width=7.0cm,height=5.0cm,angle=0]{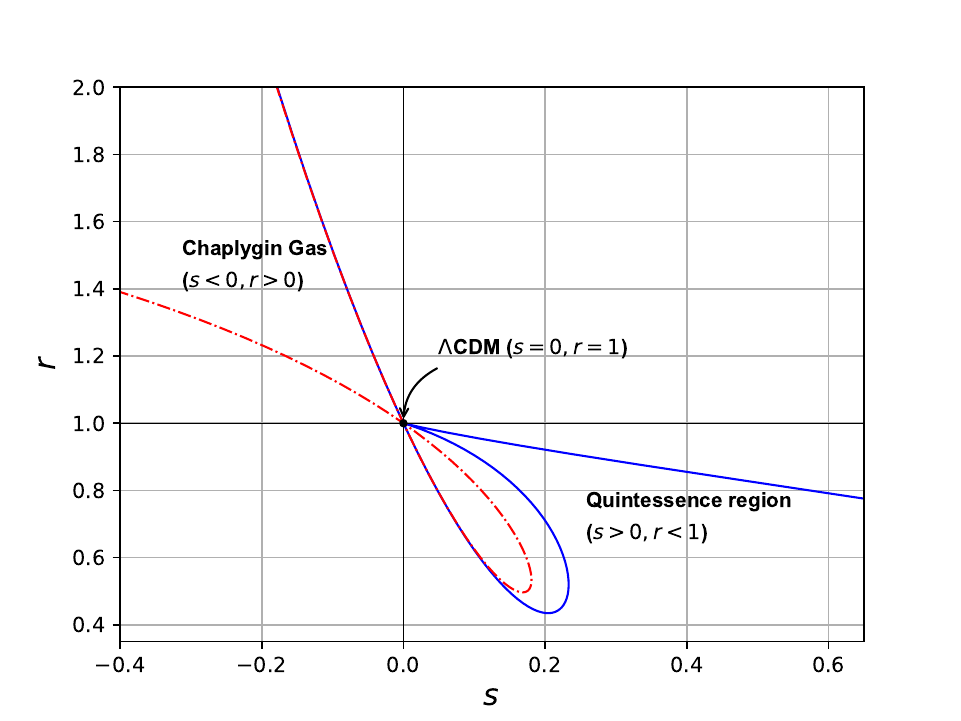}
	\caption{(a) Graphical plot of $r$ vs $q$, (b) plot of $s$ vs $q$, (c) Plot of $r$ vs $s$.}
\end{figure}
%%%%%%%%%%%%%%%%%%%%%%%%%%%%%%%%%%%%%%%%%%%%%%%%%%%%%%%%%%%%%%%%%%%%%%%%%%%%%%%%%%%%%%%%%%

\subsection{Jerk Parameter}

The jerk parameter ($j$) serves as a widely utilized tool in the field of cosmology to diagnose the various cosmological models. The concept arises from the belief that there must be a sudden jolt to shift the universe from a decelerating to an accelerating phase. The jerk is defined as the acceleration's rate of change over time, from a physical perspective. It originates from the third-order term of Taylor's series expansion of the scale factor $a$ centered at $a_0$ in cosmology. This parameter helps to differentiate between cosmological models that are kinematically degenerate. Involvement of the third order derivative of the scale factor results in increased accuracy when describing the expansion of the universe compared to the Hubble parameter.  The parameter $j$ for the suggested model can be expressed as \cite{ref77}:
\begin{eqnarray}
j&=&1-(1+z) \frac{H'(z)}{H(z)}+\frac{1}{2}(1+z)^2 \left[\frac{H''(z)}{H(z)}\right]^2,
\end{eqnarray} 
where $H'(z)$ and $H''(z)$ respectively denote the first and second order derivatives of parameter $H(z)$ w.r.t. redshift $z$. 

For the given model parameter $j$ can be recast as:
\begin{eqnarray}
j &=&\frac{1}{(4 n-2)(2 n-1)} \bigg[3 n (z+1) (2 n-1) (\alpha +2 \beta  (z+1))\nonumber\\
&+&\bigg(\left(n \left(3 \alpha  (z+1)+3 \beta  (z+1)^2-4\right)+2\right)\nonumber\\ &\times&\left(n \left(3 \alpha  (z+1)+3 \beta  (z+1)^2-2\right)+1\right)\bigg)\bigg].
\end{eqnarray}

%%%%%%%%%%%%%%%%%%%%%%%%%%%%%%%%%%%%%%% Figure 5 %%%%%%%%%%%%%%%%%%%%%%%%%%%%%%%%%%%%%%%
\begin{figure}
	\centering
	\includegraphics[scale=0.5]{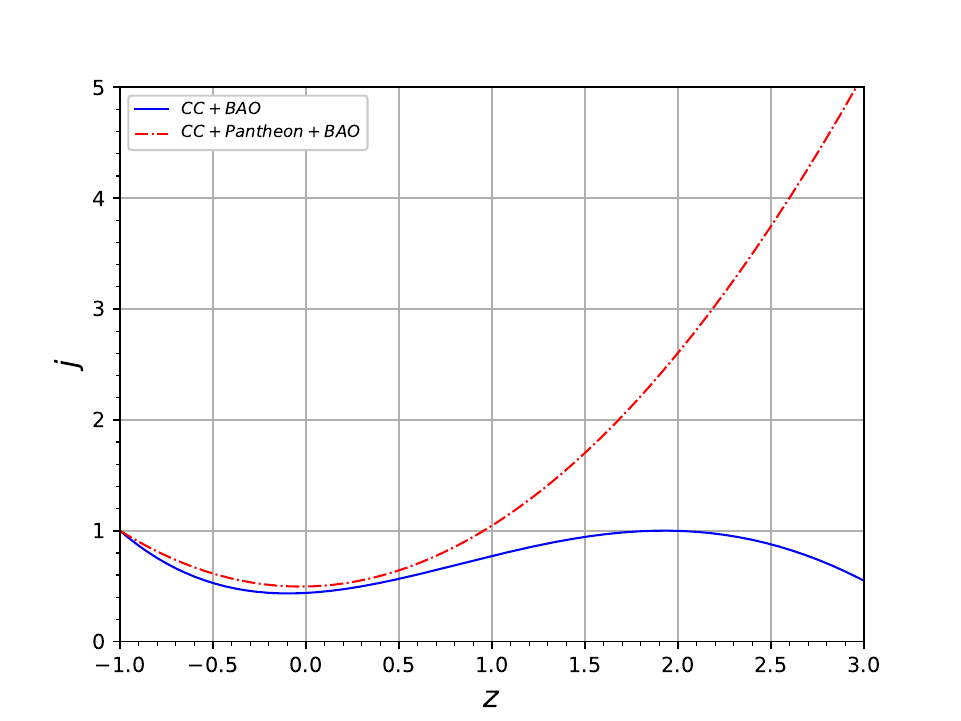}
	\caption{Behaviour of jerk parameter $ j $ with $ z $. }	
\end{figure}
%%%%%%%%%%%%%%%%%%%%%%%%%%%%%%%%%%%%%%%%%%%%%%%%%%%%%%%%%%%%%%%%%%%%%%%%%%%%%%%%%%%%%%%%

Figure 7 depicts the graphical nature of parameter $j$ as the redshift function for joint dataset of CC, BAO, and Pantheon observational sets. Earlier theoretical studies suggest that positive values of the jerk parameter indicate an accelerated expansion of the universe \cite{ref78,ref79,ref80}. Based on our model, the jerk parameter is initially negative, and then increases to eventually converge with the $\Lambda$CDM model over time. The graph demonstrates that the parameter $j$ for the given model approaches 1, aligning with findings ($q=-1$, $j=1$) as indicated by the typical $\Lambda$CDM model. In the existing analysis, the present value of the parameter $j_0$ is determined to be $0.77$ for the joint observational dataset.

\subsection{Om diagnostic}

The parameter $ O_{m}$ serves as an additional diagnostic tool in cosmology, as introduced by Sahni et al. \cite{ref81}. This parameter can be derived solely from the Hubble parameter, eliminating the need for $H'(z)$ or other pertinent data, which consequently reduces the likelihood of errors. Its ability, allowing for reconstruction through both nonparametric as well as parametric approaches, has contributed to the popularity of $O_{m}$ in the cosmological investigations. Furthermore, it is important to mention that the $O_{m}$ can differentiate between various cosmic models even in the absence of information regarding the matter density parameter and the equation of state (EOS) \cite{ref81,ref82}. According to the standard formulation \cite{ref81}, the expression for $ O_{m} (z) $ in the given model can be expressed as follows:
\begin{equation}
Om(z)=\frac{e^{\frac{3 n z \left(\alpha +\beta +\frac{\beta  z}{2}\right)}{2 n-1}}-1}{(z+1)^3-1}.
\end{equation}

%%%%%%%%%%%%%%%%%%%%%%%%%%%%%%%%%%%%%%% Figure 6 %%%%%%%%%%%%%%%%%%%%%%%%%%%%%%%%%%%%%
\begin{figure}
	\centering
	\includegraphics[scale=0.5]{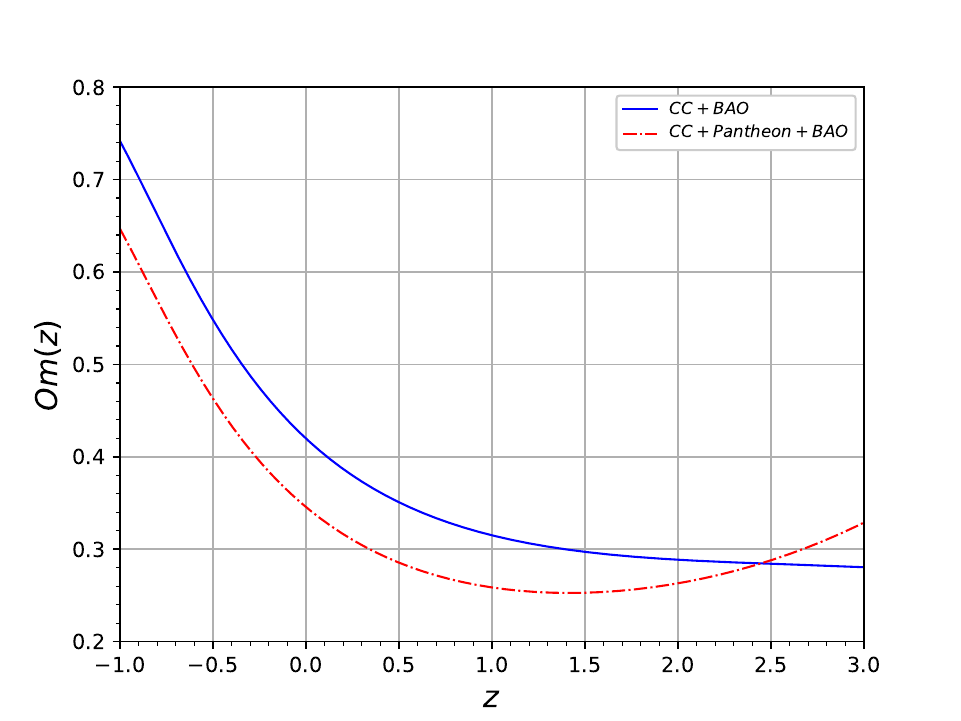}
	\caption{Evaluation of $ Om(z) $.}	
\end{figure}
%%%%%%%%%%%%%%%%%%%%%%%%%%%%%%%%%%%%%%%%%%%%%%%%%%%%%%%%%%%%%%%%%%%%%%%%%%%%%%%%%%%%%%%%%%%%%%%%%%%%%%%%%%%%%%%%%%%%%%%%% 

The evolution trajectory of $O_{m} (z)$ has been displayed in Fig. 8 using joint dataset. The incline of the diagnostic $O_{m}(z)$ serves as a crucial indicator for determining the type of dark energy (DE) cosmic models. Specifically, a negative slope suggests a quintessence-like model, while a positive slope indicates a phantom-like cosmic model. Furthermore, a zero curvature in the slope of $O_{m}(z)$ implies a $\Lambda$CDM type of DE model. The $O_{m}(z)$ analysis of our model reveals a negative slope, indicating a quintessence-like behavior, which supports the belief of a dark energy-dominated universe. The suggested cosmological model is distinguished from the standard $\Lambda$CDM model by its positive curvature \cite{ref78,ref82,ref83}.

\section{Energy Conditions} \label{sec5}

%%%%%%%%%%%%%%%%%%%%%%%%%%%%%%%%%%%%%%%%%%%%%%%%%%%%%%%%%%%%%%%%%%%%%%%%%%%%%%%%%%%%%%%%%%%%%%%%%%%%%%%%%%%%%%%%%%%%%%%%%%%%%%%%%%%%%%%%%%%%%%%%%%%%%%%%%%%%%%%%%%%%%%%
\begin{figure}
	\centering
	(a)\includegraphics[width=7cm,height=5cm,angle=0]{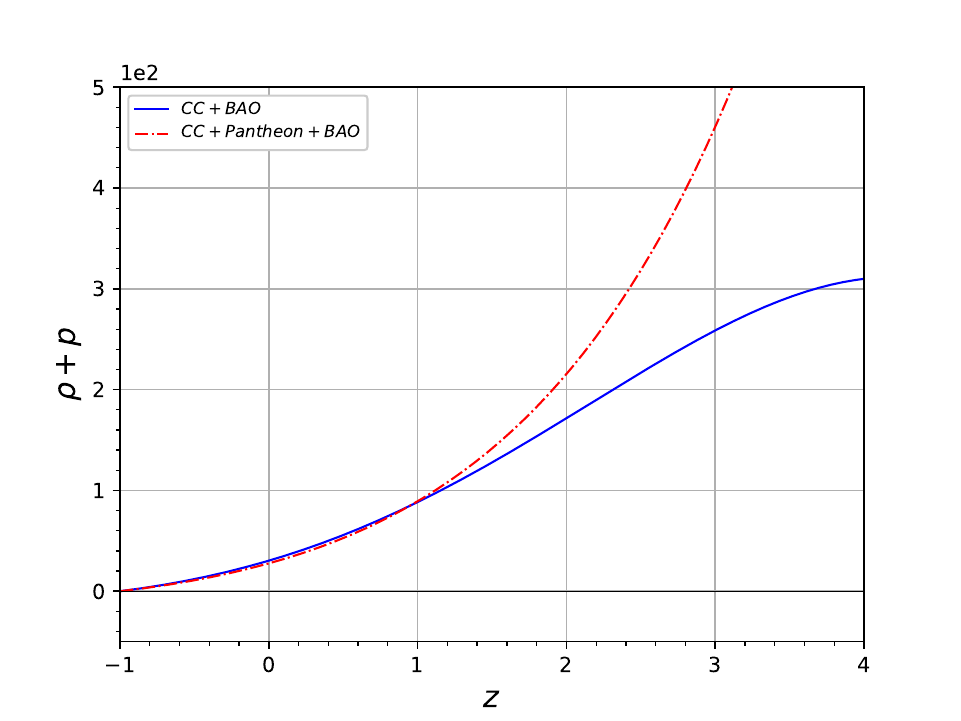}
	(b)\includegraphics[width=7cm,height=5cm,angle=0]{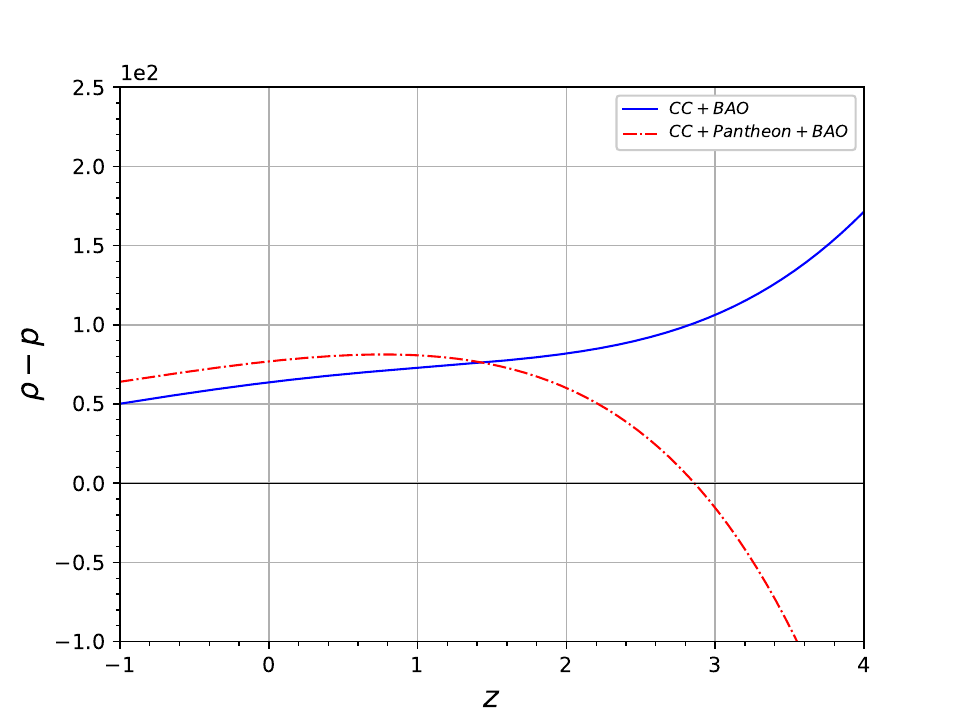}
	(c)\includegraphics[width=7cm,height=5cm,angle=0]{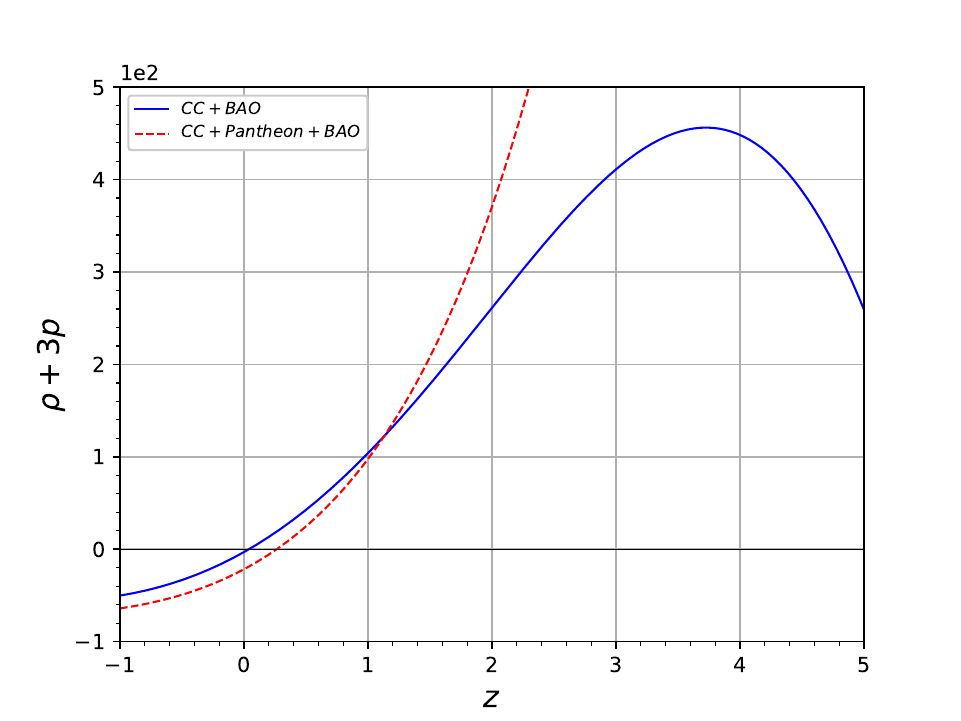}
	\caption{(a) Graphical plot of NEC, (b) plot of DEC, (c) plot of SEC.}
\end{figure}
%%%%%%%%%%%%%%%%%%%%%%%%%%%%%%%%%%%%%%%%%%%%%%%%%%%%%%%%%%%%%%%%%%%%%%%%%%%%%%%%%%%%%%%%%%%%%%%%%%%%%%%%%%%%%%%%%%%%%%%%%%%%%%%%%%%%%%%%%%%%%%%%%%%%%%%%%%%%%%%%%%%%%%%

We have discussed the physical viability of the derived model via the evolution of energy conditions (ECs) \cite{ref84,ref85}. The energy conditions for cosmic model in $f(Q)$ theory are defined as \cite{ref84,ref85,ref86}: (i) Weak energy conditions (WEC) if $\rho \geq0$, (ii) Dominant energy conditions (DEC) if $\rho - p \geq 0$, (iii) Null energy conditions (NEC) if $\rho + p \geq 0$, and (iv) Strong energy conditions (SEC) if $\rho + 3 p \geq 0$. 

Generally, the WEC and DEC are always satisfied by all recognized energy and matters \cite{ref13,ref87,ref88}. The unusual energy (DE) that creates a strong negative pressure is responsible for the universe's fast expansion and the violet SEC \cite{ref13,ref89,ref90}. Figure shows the behavior of various energy conditions in $f(R,\mathcal{L}_{m})$ gravity, utilizing a combined datasets of CC, Pantheon, and BAO. In this analysis, both the Null Energy Condition (NEC) and the Dominant Energy Condition (DEC) are satisfied while Strong Energy Condition (SEC) violates for the proposed model as depicted in Figure that validates an accelerated expansion of cosmos in the present era \cite{ref13,ref89,ref90}.

\section{Discussion and Conclusion} \label{sec6}

In this study, we have explored a transitioning cosmological model in $f(R,\mathcal{L}_{m})$ gravity. The quadratic form of equation of state parameter is considered to determine the explicit solution fo field equations in $f(R,\mathcal{L}_{m})$ gravity. The model parameters are estimated taking observational datasets of BAO, Pantheon, and CC using MCMC analysis. Some dynamical properties like deceleration parameter, jerk parameter are described. The important characteristics of the proposed model are given below:  

\begin{itemize}
\item[i)] The confidence contour plot of model parameter for joint data set of CC, BAO, and Pantheon is presented in Fig. 1. The best estimated values of the model's free parameters by employing CC, Pantheon, BAO and their combined data sets are tabulated in Table 1. At this juncture, it is to mention that we have used the GitHub repository in connection to Fig. 1. Furthermore, to constrain the model parameters, we have used the freely available GetDist python code\cite{Get}.
	
\item[ii)] The comparative behavior of Hubble rate $H(z)/(1+z)$ and standard $\Lambda$CDM is plotted in Fig. 2(a) and distance modulus $\mu(z)$ against $z$ in Fig. 2(b). From both the Figs. 2(a) and 2(b) it is clearly observed that the derived model is in nice agreement with standard $\Lambda$CDM model and follows CC and Pantheon observations with great understanding. 
	
\item[iii)]The behavior of energy density for scalar field model remains positive during entire evolution of the universe while cosmic pressure of scalar field remains negative. The natures of energy density and cosmic pressure of the proposed model for combined observational dataset can be seen in Fig. 3.
	
\item[iv)] From Fig. 4, it has been observed that the EOS parameter for derived model begins with the initial big-bag era crosses the quintessence era and approaches $\Lambda$CDM in late-time. 
	
\item[v)] The model shows a signature flipping nature of cosmos from deceleration phase to present accelerated expansion era of universe with transition redshift at $z_{t} = 0.942^{+0.112}_{-0.164}$ with the present value of DP as $q_0 = -0.4815^{+0.0362}_{-0.0096}$ as presented in Fig. 5. The current value of the deceleration parameter in the proposed model for Pantheon data is calculated as $-0.5866^{+0.1146}_{-0.1171}$ aligning well with the most recent observational findings.

\item[vi)] Figure 6 shows that the suggested model is situated within the Chaplygin gas region $(r > 1, s < 0)$ and converges to the $\Lambda$CDM point $(r = 1, s = 0)$ at the late passes of the time line. The late time acceleration is confirmed by the positive value of the jerk measure as shown in Fig. 7.
	
\item[vii)] The negative slope of $O_m(z)$ diagnostic parameter discriminates the proposed cosmic model  with standard  $\Lambda$CDM model. The result confirms the quintessence like behavior of the proposed model as clear from Fig. 8.	
	
\item[viii)] In connection to Fig. 9, the violation of SEC in late time for the derived model depicts the present expansion of universe at faster rate and indication of exotic matter in the universe. It is to be noted that energy conditions are altogether some straightforward restrictions which are normally imposed on various linear combinations of energy density and pressure so that the non-negativity of energy density and an attractive nature of gravity can be maintained as emerged from the Raychaudhuri equation~\cite{Visser2000,Carroll2004}. However, the violation of the SEC in both the present and future, indicating that the accelerated expansion of the universe is driven by a form of DE that does not contribute to the gravitational focusing of geodesics. Therefore, the violation of the SEC, along with $\omega< -1/3$, suggests the quintessence-like behavior of the universe, which is associated with its accelerated expansion at the late-time. Moreover, generally, the modified gravity theories like $f(T)$, $f(R,T)$, $f(Q)$, $f(Q,T)$, $f(R,L_m)$, $f(Q,L_m)$ etc showed the violation of SEC due to a phantom era ($omega < -1$) and also due to the late-time accelerated expansion of the universe without aid of DE (i.e. the so-called Cosmological constant of Einstein) \cite{Singh2023,R10,R8,Sharma2025}. 

\item[ix)] It is important to highlight here that a linear EOS parameter is a specific form of the EOS parameter where $\omega$ is expressed as a function of redshift ($z$) as $\omega(z) = \omega_0 + \omega_{1} z$, where $\omega_0$ and $\omega_1$ are constants. A linear form allows for a simple and relatively easy-to-model description of the evolution of the EOS parameter, which can be important for studying the behavior of dark energy and other cosmological phenomena. In cosmology, the Chevallier–Polarski–Linder (CPL) model~\cite{Linden2008,Scherrer2015,Pan2020} is a dynamical dark energy model that parametrizes the EOS of dark energy as $\omega(z) = \omega_0 + \omega_{a} z/(1 + z)$, where $\omega_0$ and $\omega_{a}$ are free parameters, allowing for a time-varying EOS. Weller and Albrecht \cite{WA} have showed that the quadratic EOS parameter provided for a best-fit compared to other parameterizations. 
	
\end{itemize}
For the proposed model, some derived results for distinct datasets are summarized in Table 2. The scenario of present accelerated expansion of the universe is described in $f(R,\mathcal{L}_{m})$ gravity. The obtained results are in excellent agreement with recent findings.

\section*{Data Availability}             
\noindent This research did not yield any new data.

\section*{Conflict of Interest} 
\noindent The authors declare no conflict of interest.

\section*{Acknowledgement}
\noindent SR would like to express his gratitude for the Visiting Research Associateship Programme at the Inter-University Centre for Astronomy and Astrophysics (IUCAA) in Pune, India and also acknowledges for the facilities at ICARD, GLA University, Mathura. \\

\end{document}